\documentclass[11pt]{article}
\usepackage{xspace}

\usepackage[final]{acl}

\usepackage{times}
\usepackage{latexsym}

\usepackage[T1]{fontenc}

\usepackage[utf8]{inputenc}

\usepackage{inconsolata}

\usepackage{graphicx}
\usepackage{enumitem}

\usepackage{kotex}
\usepackage{booktabs}
\usepackage{multirow}
\usepackage{subfig}
\usepackage{amsmath,amssymb}
\usepackage{bbm}

\usepackage[most]{tcolorbox}
\usepackage{xcolor}
\definecolor{black}{RGB}{0,0,0}
\definecolor{deepblue}{RGB}{0,0,140}
\definecolor{deepgreen}{RGB}{0,110,0}
\definecolor{deeppurple}{RGB}{90,0,130}
\definecolor{deepbrown}{RGB}{130,80,30}
\definecolor{darkorange}{RGB}{178, 80, 24}
\usepackage[table]{xcolor} 
\title{
\textbf{\textsc{ToolDF}}: Tool-Integrated Reasoning for\\
Mixed-Authenticity Audio Deepfake Detection
}

\author{
 \textbf{Taewoo Kim\textsuperscript{1,2}},
 \textbf{Young Han Lee\textsuperscript{1}},
 \textbf{Nam In Park\textsuperscript{3}},
 \textbf{Chanwoo Kim\textsuperscript{2}}\thanks{Corresponding author}
 \\
 \textsuperscript{1}Multi-Modal Research Center, KETI, South Korea \\
 \textsuperscript{2}Department of Artificial Intelligence, Korea University, South Korea \\
 \textsuperscript{3}Digital Analysis Section, National Forensic Service, South Korea \\
 \texttt{\normalsize{
 \{kimtaewoo, yhlee\}@keti.re.kr, naminpark@korea.kr, chanwcom@korea.ac.kr} 
 }
}

\begin{document}
\maketitle

\begin{abstract}
Audio deepfake detection is commonly formulated as clip-level binary classification of single-domain audio. However, real-world manipulated audio can exhibit mixed authenticity, where genuine and manipulated cues coexist across temporal transitions, overlapping sources, or both. This setting requires not only detecting manipulated audio but also localizing the components that provide evidence for the decision. We propose ToolDF, a tool-integrated reasoning framework for mixed-authenticity audio deepfake detection. ToolDF employs an audio large language model as an orchestrator trained with supervised tool-use trajectories. It adaptively analyzes the audio scene, selectively performs source separation, routes components to domain-specific experts, and aggregates their evidence into an interpretable verdict. We further introduce a mixed-authenticity ADD benchmark covering temporal transitions, acoustic overlaps, and hybrid mixtures. Experimental results show that ToolDF achieves the best overall performance on composite-type detection, achieving macro-F1 gains of 3.72 and 14.39 points over the strongest monolithic baseline and a fixed pipeline, respectively, while providing interpretable evidence localized to temporal regions and acoustic sources. Our source code and dataset are publicly available online\footnote{\url{https://github.com/rlataewoo/tooldf}}.
\end{abstract}

\section{Introduction}
\label{sec:introduction}
Audio deepfake detection (ADD) is typically formulated as clip-level binary classification of single-domain audio~\cite{jung2022aasist, tak22_odyssey, rosello23_interspeech, kim25j_interspeech}, where an input clip is classified as either real or fake. While this formulation has supported significant progress across speech, singing, music, and environmental-sound domains~\cite{zang24_interspeech, zang2024singfake, afchar2025ai, xie2025codecfake, xie2026detect}, it becomes restrictive when an audio input contains multiple acoustic components with different authenticity states.

To address this limitation, we introduce the task of mixed-authenticity audio deepfake detection, where genuine and manipulated cues coexist within a single audio input. While recent efforts have begun addressing localized manipulations or specific acoustic domains~\cite{zhang2022partialspoof, zang2024singfake, zhang2026compspoof}, real-world mixed-authenticity audio may involve complex manipulations across temporal transitions, overlapping acoustic sources, or both. For example, an audio clip might feature a transition from genuine speech to synthetic singing, layered over genuine background music. A robust detection system should therefore not only determine whether the audio is manipulated but also localize the specific temporal segments or acoustic sources that provide evidence for the final decision.

Existing ADD systems are not designed to handle such mixed-authenticity inputs. Domain-specific detectors can become unreliable when an input contains out-of-domain sources, while clip-level binary classifiers may overlook localized manipulations~\cite{li2024cross}. Fixed pipelines that rely on predefined separation or detection steps can also be suboptimal~\cite{zang2024singfake, zhang2026compspoof}, as separation can help with overlapping sources but may introduce artifacts when the input does not require separation. Moreover, using an audio large language model (ALLM)~\cite{gu2025allm4add, guo2026towards} directly as a holistic binary classifier does not leverage specialized detectors or reveal what evidence supports the prediction. These limitations motivate an adaptive framework that analyzes the input structure, selectively invokes expert tools, and integrates their outputs into an interpretable authenticity decision.

In this paper, we propose ToolDF, a tool-integrated reasoning (TIR) framework that employs an ALLM as an orchestrator rather than a direct authenticity classifier. Given an audio input, ToolDF analyzes its structure to identify relevant temporal segments and acoustic sources. It then selectively invokes domain-specific expert detectors for speech, singing, music, and environmental sound. When overlapping sources are identified, ToolDF applies source separation before localized detection. Finally, it aggregates the tool responses into an interpretable authenticity decision.

To achieve this, we train ToolDF via supervised fine-tuning (SFT) on structured tool-use trajectories. Each trajectory explicitly encodes the structured reasoning process, demonstrating how the model analyzes the input structure, plans tool calls, interprets expert responses, and derives the final verdict. Instead of relying solely on the final authenticity label, these trajectories provide rich intermediate supervision for structural analysis, tool selection, and evidence aggregation.

To evaluate the proposed framework, we construct a comprehensive mixed-authenticity audio benchmark covering both single-type and composite-type manipulation scenarios. Experimental results demonstrate that ToolDF achieves the best overall performance on composite-type detection, particularly in scenarios requiring reasoning over temporal segments and overlapping sources. Furthermore, ToolDF provides interpretable tool-execution traces, clearly revealing which segments, sources, and detector outputs contribute to the final decision.

Our contributions are summarized as follows:
\begin{itemize}
    \item We formulate the task of mixed-authenticity audio deepfake detection, where genuine and manipulated cues coexist across temporal transitions and overlapping acoustic sources.
    \item We propose ToolDF, an ALLM-based TIR framework that acts as an orchestrator to analyze audio structure, route expert tools, and aggregate evidence through supervised tool-use trajectories.
    \item We construct and publicly release a comprehensive mixed-authenticity benchmark, demonstrating through extensive experiments that ToolDF achieves the best overall performance on composite-type detection while providing transparent, interpretable component-level evidence.
\end{itemize}

\section{Related Work}
\label{sec:related_work}
\begin{figure*}[t]
    \centering
    \includegraphics[width=\textwidth]{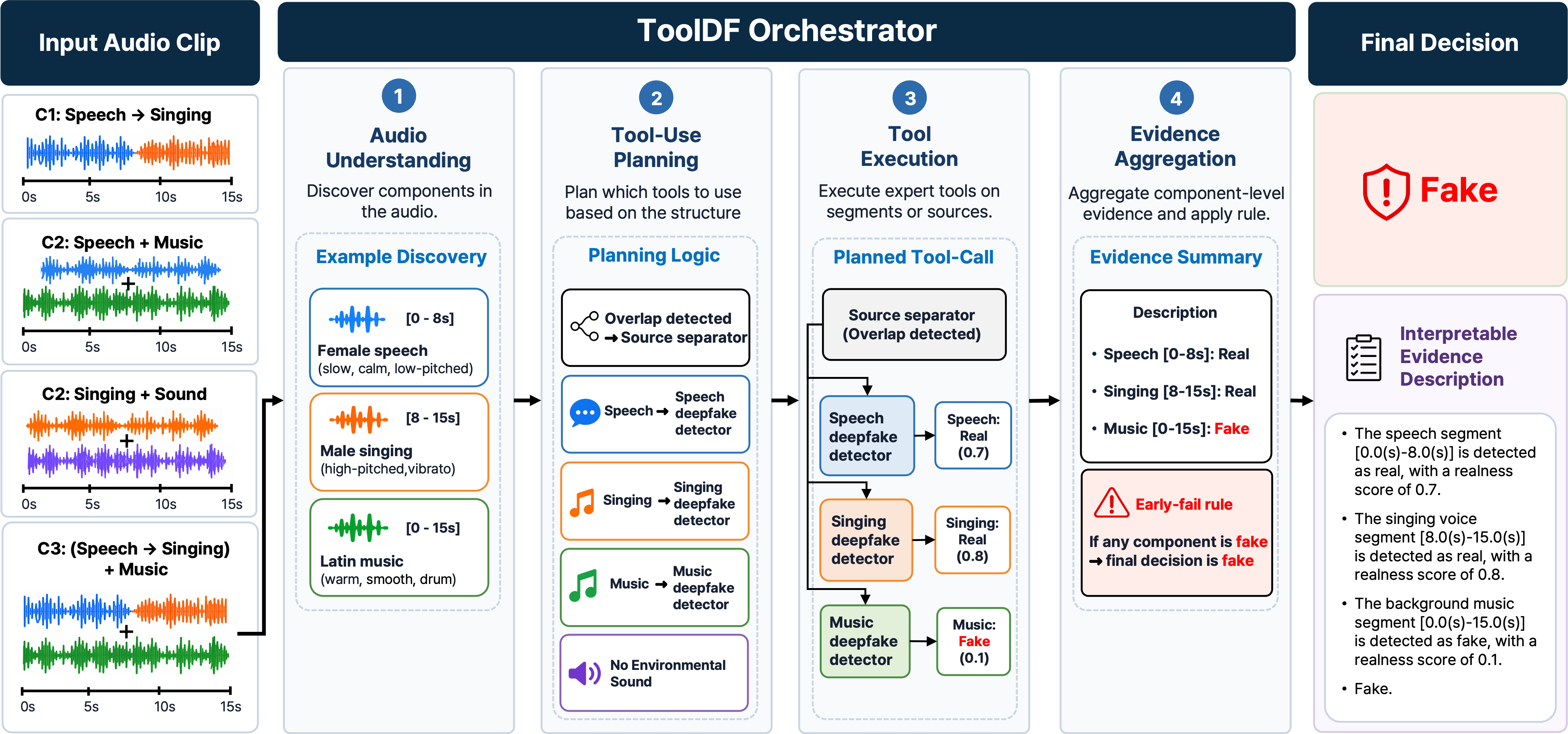}
    \caption{Overview of the ToolDF framework for mixed-authenticity audio deepfake detection.}
    \label{fig:tooldf_framework}
\end{figure*}

\subsection{Audio Deepfake Detection}
ADD has been studied extensively in the speech domain, with progress driven by benchmark datasets such as ASVspoof~\cite{yamagishi21_asvspoof} and architectures such as AASIST~\cite{jung2022aasist}. Recent efforts have expanded the scope of ADD from speech to diverse acoustic domains, including singing, music, and environmental sounds~\cite{zang24_interspeech, zang2024singfake, afchar2025ai, xie2025codecfake, xie2026detect}. However, these methods assume that each input clip belongs to a single dominant acoustic domain and predict a single authenticity label for the entire audio. This paradigm becomes restrictive when multiple audio types and varying authenticity states coexist within a single input.

\subsection{Multi-Source and Composite Audio ADD}
To address composite audio inputs, source separation has been explored to isolate source-specific manipulation cues~\cite{zang2024singfake, zhang2026compspoof}. For instance, SingFake~\cite{zang2024singfake} evaluates deepfake detection performance on both song mixtures and separated vocals, noting that separation may introduce artifacts that obscure spoofing cues. More recently, CompSpoof~\cite{zhang2026compspoof} combines mixture detection with source separation and component-level anti-spoofing; its downstream detection remains limited to speech and environmental sounds, without explicitly treating singing and music as distinct domains for detection. In contrast, \emph{ToolDF} addresses audio containing temporal transitions, acoustic overlaps, or their combinations, selectively invoking source separation when structurally required and evaluating authenticity across speech, singing, music, and environmental sounds using domain-specific detectors.

\subsection{ALLM and Tool-Integrated Reasoning}
ALLMs have demonstrated strong capabilities in understanding speech, music, and environmental sounds~\cite{rouditchenko2025omni}, leading to early explorations in audio deepfake detection~\cite{gu2025allm4add, guo2026towards, xie2026interpretable}. A straightforward approach is to deploy an ALLM directly as a holistic binary classifier. However, such direct prediction operates as a black box, providing limited insight into the underlying evidence and failing to exploit domain-specific detectors. Tool-augmented language models address these limitations by invoking external modules to perform expert tasks and integrating intermediate responses into a final output~\cite{schick2023toolformer, qin2024toolllm, qian2025toolrl}. ToolDF leverages supervised tool-use trajectories to operationalize tool-integrated reasoning (TIR) for ADD, training an ALLM to act as an orchestrator that analyzes mixed-authenticity inputs, routes components to domain-specific expert detectors, and aggregates component-level evidence through a structured reasoning process.

\section{Methodology}
\label{sec:methods}
\subsection{Task Formulation}
\label{sec:task_formulation}
Let $x$ denote an input audio clip. Unlike conventional ADD, where each input is assigned a single clip-level label, we assume that $x$ may contain multiple heterogeneous components:
\[
\mathcal{C}(x) = \{c_1, c_2, \ldots, c_K\}.
\]
Each component $c_k$ is associated with a content type $t_k \in \{\texttt{speech}, \texttt{singing}, \texttt{music}, \texttt{sound}\}$, a support region $\rho_k$, and a component-level authenticity label $y_k \in \{\texttt{real}, \texttt{fake}\}$. For temporally localized components, $\rho_k$ is represented by onset and offset times $(s_k,e_k)$; otherwise, it denotes a source-specific region in the acoustic mixture. The clip-level label $y \in \{\texttt{real}, \texttt{fake}\}$ follows an early-fail rule:
\[
y = \texttt{fake} \Longleftrightarrow \exists c_k \in \mathcal{C}(x) \;\; \text{s.t.} \;\; y_k = \texttt{fake}.
\]
The standard single-domain ADD problem is a restricted case where $K=1$ and the single component spans the entire clip. In the mixed-authenticity setting, a system should predict the clip-level label $y$ while identifying the component-level evidence that supports the decision.

\label{sec:tir_rollout}
\begin{figure*}[t]
\centering
\begin{tcolorbox}[
colback=gray!5!white,
colframe=black,
title=ToolDF Rollout,
boxrule=0.3mm,
width=0.98\textwidth,
arc=3mm,
auto outer arc=true,
fontupper=\footnotesize,
fonttitle=\small
]

\textbf{User Prompt}
\quad
\textcolor{deepblue}{\textless{}audio\textgreater{}} Is this audio real or fake?

\vspace{2.0mm}
\textbf{Audio Understanding and Planning}\\
\textcolor{deepblue}{\textless{}think\textgreater{}} \textcolor{darkorange}
{\textless{}audio\_understanding\textgreater{}}
[0.63(s)-2.34(s)] A female speech at a slow pace in a calm, low-pitched tone.
[2.43(s)-8.37(s)] A male singing voice with a clear, high-pitched tone and gentle vibrato.
[0.0(s)-8.5(s)] A continuous high-pitched electronic beep with a steady tone.
Environmental sound/noise detected throughout. \\
\textcolor{darkorange}{\textless{}/audio\_understanding\textgreater{}}\\
\textcolor{teal}{\textless{}plan\textgreater{}}\\
1. Background sound overlaps with vocals. Call \texttt{source\_separator} first.\\
2. [0.63(s)-2.34(s)] Speech detected. Call \texttt{speech\_deepfake\_detector} on the separated vocals.\\
3. [2.43(s)-8.37(s)] Singing voice detected. Call \texttt{singing\_deepfake\_detector} on the separated vocals.\\
4. [0.0(s)-8.5(s)] Background sound. Call \texttt{sound\_deepfake\_detector} on the separated background.\\
\textcolor{teal}{\textless{}/plan\textgreater{}}
\textcolor{deepblue}{\textless{}/think\textgreater{}}

\vspace{2.0mm}
\textbf{Tool Execution}\\
\textcolor{deepgreen}{\textless{}tool\_call\textgreater{}}
\{``name'': ``source\_separator'', ``parameters'': \{``audio\_path'': ``input\_audio'', ``vocals\_path'': ``vocals.wav'', ``background\_path'': ``background.wav''\}\}
\textcolor{deepgreen}{\textless{}/tool\_call\textgreater{}}
\textcolor{deepbrown}{\textless{}tool\_response\textgreater{}}
\{``model'': ``source\_separator'', ``status'': ``completed''\}
\textcolor{deepbrown}{\textless{}/tool\_response\textgreater{}}\\
\textcolor{deepgreen}{\textless{}tool\_call\textgreater{}}
\{``name'': ``speech\_deepfake\_detector'', ``parameters'': \{``start'': 0.63, ``end'': 2.34, ``vocals\_path'': ``vocals.wav''\}\}
\textcolor{deepgreen}{\textless{}/tool\_call\textgreater{}}
\textcolor{deepbrown}{\textless{}tool\_response\textgreater{}}
\{``prediction'': ``real'', ``score'': 0.99\}
\textcolor{deepbrown}{\textless{}/tool\_response\textgreater{}}\\
\textcolor{deepgreen}{\textless{}tool\_call\textgreater{}}
\{``name'': ``singing\_deepfake\_detector'', ``parameters'': \{``start'': 2.43, ``end'': 8.37, ``vocals\_path'': ``vocals.wav''\}\}
\textcolor{deepgreen}{\textless{}/tool\_call\textgreater{}}
\textcolor{deepbrown}{\textless{}tool\_response\textgreater{}}
\{``prediction'': ``fake'', ``score'': 0.29\}
\textcolor{deepbrown}{\textless{}/tool\_response\textgreater{}}\\
\textcolor{deepgreen}{\textless{}tool\_call\textgreater{}}
\{``name'': ``sound\_deepfake\_detector'', ``parameters'': \{``start'': 0.0, ``end'': 8.5, ``background\_path'': ``background.wav''\}\}
\textcolor{deepgreen}{\textless{}/tool\_call\textgreater{}}
\textcolor{deepbrown}{\textless{}tool\_response\textgreater{}}
\{``prediction'': ``real'', ``score'': 0.95\}
\textcolor{deepbrown}{\textless{}/tool\_response\textgreater{}}

\vspace{2.0mm}
\textbf{Evidence Aggregation}\\
\textcolor{deeppurple}{\textless{}description\textgreater{}}
The speech segment [0.63(s)-2.34(s)] is detected as real, with a realness score of 0.99.
The singing voice segment [2.43(s)-8.37(s)] is detected as fake, with a realness score of 0.29.
The background sound segment [0.0(s)-8.5(s)] is detected as real, with a realness score of 0.95.
\textcolor{deeppurple}{\textless{}/description\textgreater{}}\\
\textcolor{deeppurple}{\textless{}answer\textgreater{}fake\textless{}/answer\textgreater{}}

\end{tcolorbox}
\caption{Execution rollout of the structured tool-integrated reasoning framework.}
\label{fig:tir_rollout}
\end{figure*}

\subsection{ToolDF Framework}
\label{sec:framework}
ToolDF formulates mixed-authenticity audio deepfake detection as a structured TIR process rather than direct clip-level classification. Given an input audio clip $x$, an ALLM serves as an orchestrator that analyzes the audio scene, plans tool usage, invokes domain-specific expert tools, and aggregates their observations into a final authenticity decision. As illustrated in Figure~\ref{fig:tooldf_framework}, the framework consists of four phases: audio understanding, tool-use planning, localized tool execution, and evidence aggregation.

Specifically, ToolDF implements TIR as a trajectory that links audio-scene analysis with authenticity assessment. Rather than relying on unconstrained free-form rationales, ToolDF represents the reasoning process as an explicit sequence of structured decisions: identifying the constituent acoustic components $\mathcal{C}(x)$, routing each component $c_k$ to suitable expert tools based on its content type $t_k$, executing localized tool calls over the support region $\rho_k$, and deriving the final clip-level verdict $y$ from component-level evidence. Figure~\ref{fig:tir_rollout} shows an example rollout of this structured trajectory.

\paragraph{Audio Understanding.}
Given an input audio clip $x$, the orchestrator first produces a structured \texttt{<audio\_understanding>} block that summarizes the acoustic composition of the clip. The block identifies a set of components $\mathcal{C}(x)=\{c_1,\ldots,c_K\}$, where each component $c_k$ is associated with a content type $t_k \in \{\texttt{speech}, \texttt{singing}, \texttt{music}, \texttt{sound}\}$ and a support region $\rho_k$. This step helps characterize the acoustic structure of the input and informs the selection and use of downstream detectors.

\paragraph{Planning.}
Based on the identified components $\mathcal{C}(x)$, the orchestrator formulates a conditional tool-use sequence within the \texttt{<plan>} block. If vocal and background components overlap in time, the plan prioritizes the invocation of a source separator to isolate distinct acoustic streams. The resulting components are then mapped to the corresponding expert detectors according to their content types $t_k$.

\paragraph{Tool Execution.}
The orchestrator executes the plan by issuing structured tool calls. As shown by the \texttt{<tool\_call>} and \texttt{<tool\_response>} sequences in Figure~\ref{fig:tir_rollout}, each call specifies which tool to use and what input region to analyze, such as a temporal segment or a separated source. The called tool then returns a component-level authenticity prediction $\hat{y}_k$ for the target component. These tool outputs provide external evidence for the final decision.

\paragraph{Evidence Aggregation.}
After receiving the tool observations, the orchestrator summarizes the component-level evidence within the \texttt{<description>} block and derives the final clip-level decision. Following the mixed-authenticity formulation in Section~\ref{sec:task_formulation}, the final output in the \texttt{<answer>} tag is \texttt{fake} if any component-level prediction $\hat{y}_k$ is \texttt{fake}. Otherwise, the framework returns \texttt{real}.

\begin{table*}[t]
\centering
\small
\setlength{\tabcolsep}{4pt}
\renewcommand{\arraystretch}{0.89}
\begin{tabular*}{\textwidth}{@{\extracolsep{\fill}}llrrr}
\toprule
\textbf{Type} & \textbf{Acoustic Configurations} & \textbf{\# Train} & \textbf{\# Dev} & \textbf{\# Eval} \\ 
\midrule
\textbf{ASVspoof2019 LA}~\cite{nautsch2021asvspoof} & Speech & 25,380 & 24,844 & 71,237 \\
\textbf{CtrSVDD}~\cite{zang24_interspeech} & Singing Voice & 84,404 & 43,625 & 92,769 \\
\textbf{EnvSDD}~\cite{yin25_interspeech} & Environmental Sound & 139,055 & 39,710 & 39,768 \\
\textbf{FakeMusicCaps}~\cite{comanducci2025fakemusiccaps} & Music (Fake) & 16,055 & 5,350 & 6,200 \\
\textbf{MusicCaps}~\cite{agostinelli2023musiclm} & Music (Real) & 3,211  & 1,070 & 1,071  \\
\midrule
\textbf{C1} & Speech $\rightarrow$ Singing & 12,690 & 12,422 & 35,616 \\
\textit{(Temporal Transition)} & Singing $\rightarrow$ Speech & 12,690 & 12,422 & 35,620 \\ 
\midrule
\textbf{C2} & Speech + Music & 3,170 & 2,325 & 3,114 \\
\textit{(Acoustic Overlap)} & Singing + Music & 10,281 & 4,095 & 4,157 \\
                            & Speech + Environmental Sound & 22,210 & 14,060 & 17,234 \\
                            & Singing + Environmental Sound & 74,122 & 25,650 & 22,534 \\ 
\midrule
\textbf{C3}                     & (Speech $\rightarrow$ Singing) + Music & 1,418 & 1,197 & 979 \\
\textit{(Transition + Overlap)} & (Singing $\rightarrow$ Speech) + Music & 1,491 & 1,222 & 985 \\
                                & (Speech $\rightarrow$ Singing) + EnvSound & 11,272 & 7,815 & 5,061 \\
                                & (Singing $\rightarrow$ Speech) + EnvSound & 11,199 & 7,790 & 5,059 \\
\midrule
\textbf{Total} & All Configurations & 428,648 & 203,597 & 341,404 \\ 
\bottomrule
\end{tabular*}
\vspace{-2mm}
\caption{Statistics of the mixed-authenticity audio deepfake detection benchmark, including both single-type corpora and constructed composite-type configurations. The composite-type setting consists of \textbf{C1} temporal speech--singing transitions, \textbf{C2} acoustic overlaps between foreground and background sources, and \textbf{C3} hybrid mixtures combining transitions and overlaps. Details of the data construction are provided in Appendix~\ref{sec:appendix_dataset}.}
\label{tab:dataset_stats}
\vspace{-1.5mm}
\end{table*}

\subsection{Supervised Trajectory Learning}
\label{sec:sft}
We optimize ToolDF with supervised fine-tuning (SFT) to train the ALLM to follow the structured TIR protocol. Given an input audio clip $x$, the orchestrator is trained to autoregressively generate a structured trajectory from audio understanding to the final decision: \[o = \langle u, p, \mathcal{T}, d, y \rangle,\]
where $u$ denotes the audio-understanding block, $p$ the tool-use plan, $\mathcal{T}=((a_i,r_i))_{i=1}^{N}$ the sequence of tool calls $a_i$ and tool observations $r_i$, $d$ the evidence summary, and $y \in \{\texttt{real}, \texttt{fake}\}$ the final clip-level verdict.

To construct supervised trajectories, we use ground-truth component annotations:
\[
\mathcal{S}^{\star} = \{(\rho_k, t_k, y_k)\}_{k=1}^{K},
\]
where $\rho_k$, $t_k$, and $y_k$ follow the definitions in Section~\ref{sec:task_formulation}. These annotations are used to construct audio-understanding blocks, tool-use plans, and localized tool calls. During trajectory construction, the tool observations are populated using the ground-truth component authenticity labels and inserted into the trajectory as conditioning context for subsequent generation. The evidence summary and final verdict are then constructed from these ground-truth observations according to the mixed-authenticity early-fail rule.

Although the trajectory contains both generated tokens and tool observation tokens, the model is optimized only on tokens produced by the orchestrator. Let $\mathcal{M}$ denote the set of token positions corresponding to $u$, $p$, $\{a_i\}_{i=1}^{N}$, $d$, and $y$, excluding the observation tokens $\{r_i\}_{i=1}^{N}$. The SFT objective is:
\[
\mathcal{L}_{\mathrm{SFT}}
=
-\sum_{m \in \mathcal{M}}
\log P_{\theta}
\left(
o_m \mid o_{<m}, x, q
\right).
\]
Here, $x$ denotes the input audio clip, $q$ denotes the textual instruction context, $o_m$ denotes the $m$-th token in the target trajectory, $o_{<m}$ denotes the preceding tokens, and $P_{\theta}$ denotes the next-token distribution of the ALLM.

\section{Mixed-Authenticity ADD Benchmark}
\label{sec:benchmark}
To evaluate ToolDF beyond isolated single-domain settings, we construct a mixed-authenticity ADD benchmark by combining public audio deepfake datasets from multiple acoustic domains. Existing benchmarks typically focus on a single content type, such as speech, singing, music, or environmental sound, and thus do not sufficiently evaluate performance on heterogeneous audio scenes containing multiple components with different authenticity cues.

The benchmark is built from authentic and synthetic subsets of domain-specific corpora, including ASVspoof2019~\cite{nautsch2021asvspoof} for speech, CtrSVDD~\cite{zang24_interspeech} for singing, FakeMusicCaps~\cite{comanducci2025fakemusiccaps} and MusicCaps~\cite{agostinelli2023musiclm} for music, and EnvSDD~\cite{yin25_interspeech} for environmental sounds. Using these sources, we construct both single-type and composite-type evaluation settings. The single-type setting contains isolated speech, singing, music, and environmental sound clips following conventional ADD evaluation. The composite-type setting consists of three splits: \textbf{C1}, temporal transitions between speech and singing; \textbf{C2}, acoustic overlaps between foreground vocal content and background music or environmental sound; and \textbf{C3}, hybrid mixtures combining both temporal transitions and acoustic overlaps.

\section{Experiments}
\subsection{Dataset}
\label{sec:dataset}
Table~\ref{tab:dataset_stats} summarizes the statistics of our mixed-authenticity ADD benchmark. We evaluate ToolDF under both conventional single-type and the proposed composite-type settings. For single-type evaluation, we follow the official training, development, and test splits of each source corpus. As an exception, due to the unavailability of certain original audio files in MusicCaps~\cite{agostinelli2023musiclm}, a small number of real samples could not be retrieved during data collection. To prevent data leakage, their corresponding fake pairs were assigned exclusively to the test set. The remaining accessible samples from MusicCaps and FakeMusicCaps~\cite{comanducci2025fakemusiccaps} were randomly partitioned into training, development, and test sets with a $6:2:2$ ratio. Detailed configurations for the composite-type settings are provided in Appendix~\ref{sec:appendix_dataset}.

\subsection{Implementation Details}
\label{sec:implementation}
\paragraph{ALLM Backbone.}
We implement ToolDF with Qwen2.5-Omni-3B~\cite{Qwen2.5-Omni} as the ALLM backbone and fine-tune it on supervised TIR trajectories. We use Low-Rank Adaptation (LoRA) for parameter-efficient tuning, keeping the base model frozen and inserting adapters into all linear layers with rank $r=64$ and scaling factor $\alpha=16$. The model is fine-tuned using the task-specific system prompt designed to enforce the structured TIR protocol, the exact text of which is detailed in Appendix~\ref{sec:appendix_system_prompt}. Training uses 428,648 examples comprising both single-type and composite-type data. We train for three epochs with AdamW, using a learning rate of $1\times10^{-5}$, weight decay of $0.1$, bfloat16 precision, gradient checkpointing, DeepSpeed ZeRO-2, and a maximum sequence length of 4096. Training is conducted on 8 NVIDIA A40 GPUs with a batch size of 4 per GPU and gradient accumulation over 4 steps, yielding a global batch size of 128.

\paragraph{Tools.}
ToolDF uses external tools for source separation and domain-specific deepfake detection. For overlapping vocal-background mixtures, we employ Demucs v4~\cite{rouard2023hybrid} to obtain vocal and non-vocal streams. For authenticity assessment, we instantiate XLSR-AASIST expert detectors for speech, singing, music, and environmental sound, each trained on data from its corresponding single-type domain. To prevent information loss at temporal transitions, the XLSR-AASIST expert detectors are trained on the full audio duration, with audio sequences dynamically repeated to match the maximum sequence length within each batch. We also train separation-aware variants to handle separated inputs and potential separation artifacts.

At inference time, the orchestrator routes the target audio segment or separated source to the selected expert according to the generated tool-use plan. Each detector returns a binary authenticity prediction and a normalized confidence score in $[0,1]$, which are inserted into the trajectory as tool observations. The decision threshold for each detector is determined by the equal error rate on its development set. 

\begin{table*}[t]
\centering
\small
\resizebox{1.0\textwidth}{!}{%
\begin{tabular}{l|cccc|c|ccc|c}
\toprule
 \textbf{Model} & \multicolumn{5}{c|}{\textbf{Single-Type}}
 & \multicolumn{4}{c}{\textbf{Composite-Type}} \\
\cmidrule(lr){2-6} \cmidrule(lr){7-10}
 & \textbf{Speech} & \textbf{Singing} & \textbf{Sound} & \textbf{Music}
 & \textbf{S-Avg.}
 & \textbf{C1} & \textbf{C2} & \textbf{C3}
 & \textbf{C-Avg.} \\
\midrule
XLSR-AASIST \cite{tak22_odyssey} & 89.60 & 84.82 & \textbf{76.67} & 95.22 & \textbf{86.57} & \textbf{79.54} & 51.95 & 42.01 & \textbf{57.83} \\
XLSR-Conformer \cite{rosello23_interspeech} & 90.17 & 83.34 & 72.56 & 92.83 & 84.73 & 74.29 & 48.82 & 38.12 & 53.74 \\
WPT-XLSR-AASIST \cite{xie2026detect} & 90.16 & \textbf{86.91} & 69.26 & 82.45 & 82.20 & 68.41 & 47.87 & 31.46 & 49.25 \\
ALLM4ADD \cite{gu2025allm4add} & \textbf{97.64} & 61.56 & 75.54 & \textbf{95.62} & 82.56 & 62.62 & \textbf{55.51} & \textbf{53.22} & 57.12 \\
\bottomrule
\end{tabular}
}
\caption{Performance of models trained on single-type data across single-type and composite-type mixed-authenticity settings (Macro-F1, \%). S-Avg. and C-Avg. denote the average performance over single-type and composite-type settings, respectively.}
\label{tab:baseline_generalization}
\end{table*}

\begin{table*}[tb!]
    \centering
    \scriptsize
    \setlength{\tabcolsep}{4pt}
    \resizebox{\textwidth}{!}{%
    \begin{tabular}{l|c|cccc|c|ccc|c}
        \toprule
        \textbf{Model}
        & \textbf{Interp.}
        & \multicolumn{5}{c|}{\textbf{Single-Type}}
        & \multicolumn{4}{c}{\textbf{Composite-Type}} \\
        \cmidrule(lr){3-7}
        \cmidrule(lr){8-11}
        & 
        & \textbf{Speech}
        & \textbf{Singing}
        & \textbf{Sound}
        & \textbf{Music}
        & \textbf{S-Avg.}
        & \textbf{C1}
        & \textbf{C2}
        & \textbf{C3}
        & \textbf{C-Avg.} \\
        \midrule

        \rowcolor{blue!10}
        \multicolumn{11}{l}{\textit{\textbf{End-to-End Monolithic Models}}} \\
        \addlinespace[0.3em]

        XLSR-AASIST \cite{tak22_odyssey}
        & $\times$
        & 88.40
        & \textbf{84.91}
        & 69.37
        & 92.42
        & 83.78
        & 89.79
        & 73.16
        & 71.57
        & 78.17 \\

        XLSR-Conformer \cite{rosello23_interspeech}
        & $\times$
        & 88.09
        & 82.55
        & 71.80
        & 90.80
        & 83.31
        & 87.28
        & 74.46
        & 71.61
        & 77.78 \\

        WPT-XLSR-AASIST \cite{xie2026detect}
        & $\times$
        & 83.55
        & 78.81
        & \textbf{78.04}
        & 87.52
        & 81.98
        & 85.00
        & 73.73
        & 69.87
        & 76.20 \\

        ALLM4ADD \cite{gu2025allm4add}
        & $\times$
        & \textbf{98.17}
        & 60.94
        & 75.00
        & \textbf{95.20}
        & 82.33
        & 72.77
        & 64.92
        & 57.43
        & 65.04 \\

        \cmidrule{1-11}

        \rowcolor{blue!10}
        \multicolumn{11}{l}{\textit{\textbf{Component-Level Frameworks}}} \\
        \addlinespace[0.3em]

        Fixed Pipeline
        & \checkmark\textsuperscript{*}
        & 63.72
        & 55.58
        & 52.91
        & 66.25
        & 59.62
        & 55.57
        & \textbf{78.55}
        & 68.39
        & 67.50 \\

        ToolDF (Proposed) 
        & \checkmark
        & 95.78
        & 81.58
        & 76.13
        & 92.67
        & \textbf{86.54}
        & \textbf{91.21}
        & 77.66
        & \textbf{76.81}
        & \textbf{81.89} \\

        \midrule

        ToolDF (Oracle)
        & -
        & 95.96
        & 78.94
        & 76.10
        & 93.09
        & 86.02
        & 91.77
        & 79.34
        & 77.45
        & 82.85 \\

        \bottomrule
    \end{tabular}%
    }
    \caption{Performance comparison for single-type and composite-type audio deepfake detection. Values are macro-F1 scores (\%). S-Avg. and C-Avg. denote averages over single-type and composite-type settings, respectively. $\checkmark^*$~indicates limited or rule-based interpretability.}
    \label{tab:main_results}
    \vspace{-8pt}
\end{table*}

\subsection{Baselines}
To evaluate ToolDF in mixed-authenticity audio deepfake detection, we compare its performance against five baseline configurations. We categorize the evaluated systems into end-to-end monolithic models and component-level frameworks. The end-to-end monolithic baselines include \textbf{XLSR-AASIST}~\cite{tak22_odyssey}, a standard framework combining a self-supervised wav2vec 2.0~\cite{babu22_interspeech} backbone with a graph neural network, and \textbf{XLSR-Conformer}~\cite{rosello23_interspeech}, which leverages a Conformer backbone to capture both local and global temporal dependencies. We also compare against \textbf{WPT-XLSR-AASIST}~\cite{xie2026detect}, a frequency-sensitive approach proposed for all-type audio deepfake detection, and \textbf{ALLM4ADD}~\cite{gu2025allm4add}, a recent ALLM fine-tuned for audio deepfake detection.

As a component-level baseline, the \textbf{Fixed Pipeline} applies source separation via Demucs v4 to every input, routing the isolated foreground and background streams to specialized \textit{vocal} (speech and singing) and \textit{non-vocal} (music and environmental sound) XLSR-AASIST expert detectors. Both detectors are trained as separation-aware variants. The final verdict is determined by an early-fail rule, classifying the clip as fake if either detector predicts a synthetic origin.

\subsection{Evaluation Metric}
We use the macro-averaged F1 score as the clip-level evaluation metric for both single-type and composite-type settings. Macro-F1 computes F1 separately for the real and fake classes and averages the two scores equally, preventing the majority class from dominating the overall evaluation.

\begin{table*}[ht]
\centering
\small
\begin{tabular*}{\textwidth}{@{\extracolsep{\fill}}l|ccc|ccc|ccc}
\toprule
 & \multicolumn{3}{c|}{\textbf{C1 (Temporal)}} 
 & \multicolumn{3}{c|}{\textbf{C2 (Overlap)}} 
 & \multicolumn{3}{c|}{\textbf{C3 (Temporal + Overlap)}} \\ 
\cmidrule(lr){2-4} \cmidrule(lr){5-7} \cmidrule(lr){8-10}
\textbf{Configuration} 
& \textbf{PR} & \textbf{F1$_\text{parsed}$} & \textbf{F1$_\text{strict}$}
& \textbf{PR} & \textbf{F1$_\text{parsed}$} & \textbf{F1$_\text{strict}$}
& \textbf{PR} & \textbf{F1$_\text{parsed}$} & \textbf{F1$_\text{strict}$} \\
\midrule
\textbf{ToolDF} 
& \textbf{100.00} & \textbf{91.21} & \textbf{91.21} 
& 99.99 & \textbf{77.67} & \textbf{77.66} 
& 99.98 & \textbf{76.83} & \textbf{76.81} \\

\midrule
w/o Audio Understanding 
& 82.59 & 60.12 & 48.25 
& 79.85 & 48.82 & 39.77 
& 83.15 & 53.21 & 44.31 \\

w/o Planning            
& 96.71 & 45.47 & 44.08
& 84.34 & 24.57 & 20.87 
& 97.52 & 35.18 & 34.33 \\

w/o Tool Calling        
& \textbf{100.00} & 75.75 & 75.75 
& \textbf{100.00} & 66.49 & 66.49 
& \textbf{100.00} & 63.62 & 63.62 \\

w/o Description         
& 69.90 & 85.06 & 48.57 
& 23.69 & 69.77 & 18.57 
& 88.87 & 74.43 & 63.62 \\
\bottomrule
\end{tabular*}
\vspace{-2mm}
\caption{Ablation study under parsed-only and strict evaluation. PR denotes the parsing rate (\%), representing the percentage of outputs that conform to the designated trajectory format. F1$_\text{parsed}$ (\%) is computed only on these successfully parsed outputs, while F1$_\text{strict}$ (\%) is evaluated over the entire test set by treating unparsable outputs as incorrect predictions. Each ablation removes the corresponding stage from the structured ToolDF trajectory.}
\label{tab:ablation_study}
\vspace{-2mm}
\end{table*}

\vspace{-1mm}
\begin{table}[t]
\centering
\resizebox{\columnwidth}{!}{
\begin{tabular}{l|c|c|c|c}
\toprule
\textbf{Type} & \textbf{Seg-F1$_\text{mi.}$} & \textbf{Seg-F1$_\text{ma.}$} & \textbf{Event-F1$_\text{mi.}$} & \textbf{Event-F1$_\text{ma.}$} \\
\midrule
Speech & 97.76 & 98.18 & 93.42 & 93.64 \\
Singing & 95.58 & 97.74 & 95.19 & 97.19 \\
Sound & 99.95 & 99.97 & 99.92 & 99.96 \\
Music & 99.73 & 99.89 & 99.66 & 99.94 \\
C1 & 95.63 & 97.34 & 93.34 & 94.24 \\
C2 & 96.70 & 96.11 & 92.77 & 91.96 \\
C3 & 95.26 & 93.63 & 86.44 & 87.62 \\
\bottomrule
\end{tabular}
}
\vspace{-2mm}
\caption{Fine-grained localization performance of ToolDF across evaluation subsets. Segment-level and event-level F1 scores (\%) are reported with micro and macro averaging.}
\label{tab:segment_performance}
\vspace{-4mm}
\end{table}

\section{Results}
\subsection{The Single-Type Generalization Gap}
\label{sec:single_type_generalization}
Table~\ref{tab:baseline_generalization} examines whether existing models trained only on single-type data can generalize to composite mixed-authenticity inputs. While the baselines perform strongly on single-type cases, with XLSR-AASIST achieving the highest S-Avg. of 86.57, their performance drops sharply on composite inputs. The best C-Avg. falls to 57.83, again from XLSR-AASIST, with particularly large degradation in the more complex composite settings. ALLM4ADD performs comparatively well on overlapping scenarios, achieving the best scores on C2 and C3 among the evaluated baselines, although its composite-type performance remains substantially below its single-type performance.

These results show that strong performance on isolated acoustic domains does not ensure generalization to mixed-authenticity ADD. Since effectively handling composite scenarios requires reasoning over multiple temporal regions and acoustic sources rather than relying solely on monolithic clip-level classification, this generalization gap motivates ToolDF, which decomposes the acoustic scene through structured, tool-integrated reasoning.

\subsection{Evaluation on Mixed-Authenticity ADD}
\label{sec:main_results}

Table~\ref{tab:main_results} compares ToolDF with end-to-end monolithic models and component-level frameworks under a joint training setting. ToolDF achieves the best overall performance on composite inputs, yielding the highest C-Avg. of 81.89 and outperforming the strongest monolithic baseline, XLSR-AASIST, by 3.72 points. ToolDF performs particularly strongly on temporal transitions (C1) and hybrid mixtures (C3), achieving 91.21 and 76.81, respectively. Although ALLM4ADD achieves strong performance on single-type settings, its C-Avg. drops to 65.04 on composite inputs, indicating that strong single-type performance does not necessarily translate to mixed-authenticity detection, even with joint training.

Furthermore, ToolDF outperforms the Fixed Pipeline, achieving a C-Avg. score of 81.89 compared to the baseline's 67.50. While the Fixed Pipeline scores slightly higher on acoustic overlaps (C2) with 78.55 due to its mandatory source separation, its performance is substantially lower on C1 and C3, where separation may be unnecessary or introduce artifacts. ToolDF mitigates this limitation by adaptively deciding when to perform source separation. Crucially, ToolDF provides high interpretability through explicit tool-execution trajectories, whereas monolithic models do not provide explicit component-level evidence. Finally, ToolDF closely approaches its Oracle variant with a marginal gap of only 0.96 points, suggesting that the orchestrator can closely approximate oracle routing without access to ground-truth annotations.

\subsection{Ablation and Localization Analysis}
\label{sec:ablation}

Table~\ref{tab:ablation_study} reports the ablation results for composite-type inputs under both parsed-only and strict evaluation metrics. Each ablation variant is separately trained and evaluated with the corresponding trajectory stage removed. In \textit{w/o Audio Understanding}, the model generates the tool-use plan directly from the input audio and instruction without an explicit structured acoustic decomposition block; in \textit{w/o Planning}, tool calls are generated directly after audio understanding; in \textit{w/o Tool Calling}, the model produces the verdict without external detector responses; and in \textit{w/o Description}, the final answer is generated directly from the tool responses.

The full ToolDF model achieves the highest performance across all configurations while maintaining a parsing rate (PR) close to 100\%. This demonstrates that our structured trajectory design generates stable, rule-compliant reasoning paths without sacrificing detection performance. Among the individual components, removing the planning stage causes the largest drop in strict F1 scores across all splits. Eliminating audio understanding also substantially lowers performance, indicating that direct tool calling alone is insufficient without explicit audio-scene understanding. Furthermore, excluding description evidence severely degrades the parsing rate on C2 to 23.69, highlighting the importance of structured evidence aggregation for maintaining stable LLM reasoning in overlapping environments.

To evaluate fine-grained localization, Table~\ref{tab:segment_performance} assesses the localization performance of ToolDF following standard DCASE evaluation metrics~\cite{turpault2019sound}. The framework demonstrates stable localization capabilities across all subsets, yielding macro-F1 scores between 93.64 and 99.97 in single-type domains for both segment-level and event-level metrics. In composite-type settings, ToolDF maintains consistent localization performance, achieving event-level macro-F1 scores of 94.24 on temporal transitions (C1) and 87.62 on hybrid mixtures (C3). These results demonstrate that ToolDF can go beyond clip-level binary classification by localizing temporal regions and acoustic sources that provide evidence of manipulation.

\section{Conclusion}

We presented ToolDF, a tool-integrated reasoning (TIR) framework for mixed-authenticity audio deepfake detection. By leveraging an audio large language model as an orchestrator, ToolDF analyzes audio structures, adaptively routes heterogeneous components to specialized expert detectors, and aggregates component-level evidence into an interpretable clip-level verdict. Experiments on our mixed-authenticity audio deepfake detection benchmark demonstrate that ToolDF achieves the best overall performance on composite-type detection while providing fine-grained localization of the temporal regions and acoustic sources that support its decisions. These results highlight the potential of TIR as a general framework for interpretable and adaptive audio deepfake detection in complex acoustic scenes.

\section*{Limitations}

ToolDF depends on external tools, including source separation modules and domain-specific deepfake detectors. Therefore, errors from source separation or detector misclassification can propagate to the final decision, and the overall performance is bounded by the reliability and coverage of the underlying expert models.

In addition, our benchmark is constructed by composing existing public datasets. While this enables controlled evaluation with component-level annotations, it may not fully reflect the diversity, editing artifacts, source interactions, and distributional complexity of real-world manipulated media. The composition rules used for temporal transitions and acoustic overlaps may also be simpler than those encountered in real-world or adversarially edited audio.

Our evaluation does not include conventional partial-spoof benchmarks such as PartialSpoof~\cite{zhang2022partialspoof}, which focus on localized manipulations within a single speech utterance. Extending ToolDF to this setting by incorporating a partial-spoof detector as a speech expert remains an important direction for future work.

\section*{Acknowledgements}
This work was partly supported by the Institute of Information \& Communications Technology Planning \& Evaluation (IITP) grants funded by the Korea government (MSIT): (No. RS-2026-25525363, Agentic Experts for Generative-AI Inspection Solution, Contribution: 50\%), (No. RS-2025-02304828, Artificial Intelligence Star Fellowship Support Program to Nurture the Best AI Talent, Contribution: 45\%), and (No. RS-2019-II190079, Artificial Intelligence Graduate School Program at Korea University, Contribution: 5\%).

\bibliography{custom}

@inproceedings{afchar2025ai,
  title={Ai-generated music detection and its challenges},
  author={Afchar, Darius and Meseguer-Brocal, Gabriel and Hennequin, Romain},
  booktitle={ICASSP 2025-2025 IEEE International Conference on Acoustics, Speech and Signal Processing (ICASSP)},
  pages={1--5},
  year={2025},
  organization={IEEE}
}

@inproceedings{zang24_interspeech,
  title     = {{CtrSVDD: A Benchmark Dataset and Baseline Analysis for Controlled Singing Voice Deepfake Detection}},
  author    = {Yongyi Zang and Jiatong Shi and You Zhang and Ryuichi Yamamoto and Jionghao Han and Yuxun Tang and Shengyuan Xu and Wenxiao Zhao and Jing Guo and Tomoki Toda and Zhiyao Duan},
  year      = {2024},
  booktitle = {{Interspeech 2024}},
  pages     = {4783--4787},
  doi       = {10.21437/Interspeech.2024-2242},
  issn      = {2958-1796},
}

@article{xie2025codecfake,
  title={The codecfake dataset and countermeasures for the universally detection of deepfake audio},
  author={Xie, Yuankun and Lu, Yi and Fu, Ruibo and Wen, Zhengqi and Wang, Zhiyong and Tao, Jianhua and Qi, Xin and Wang, Xiaopeng and Liu, Yukun and Cheng, Haonan and Ye, Long and Sun, Yi},
  journal={IEEE Transactions on Audio, Speech and Language Processing},
  volume={33},
  pages={386--400},
  year={2025},
  publisher={IEEE}
}

@inproceedings{xie2026detect,
  title={Detect all-type deepfake audio: Wavelet prompt tuning for enhanced auditory perception},
  author={Xie, Yuankun and Fu, Ruibo and Wang, Xiaopeng and Wang, Zhiyong and Cao, Songjun and Ma, Long and Cheng, Haonan and Ye, Long},
  booktitle={Proceedings of the AAAI Conference on Artificial Intelligence},
  volume={40},
  pages={35922--35930},
  year={2026}
}

@inproceedings{jung2022aasist,
  title={Aasist: Audio anti-spoofing using integrated spectro-temporal graph attention networks},
  author={Jung, Jee-weon and Heo, Hee-Soo and Tak, Hemlata and Shim, Hye-jin and Chung, Joon Son and Lee, Bong-Jin and Yu, Ha-Jin and Evans, Nicholas},
  booktitle={ICASSP 2022-2022 IEEE international conference on acoustics, speech and signal processing (ICASSP)},
  pages={6367--6371},
  year={2022},
  organization={IEEE}
}

@inproceedings{tak22_odyssey,
  title     = {Automatic Speaker Verification Spoofing and Deepfake Detection Using Wav2vec 2.0 and Data Augmentation},
  author    = {Hemlata Tak and Massimiliano Todisco and Xin Wang and Jee-weon Jung and Junichi Yamagishi and Nicholas Evans},
  year      = {2022},
  booktitle = {The Speaker and Language Recognition Workshop (Odyssey 2022)},
  pages     = {112--119},
  doi       = {10.21437/Odyssey.2022-16},
}

@inproceedings{rosello23_interspeech,
  title     = {A conformer-based classifier for variable-length utterance processing in anti-spoofing},
  author    = {Eros Rosello and Alejandro Gomez-Alanis and Angel M. Gomez and Antonio Peinado},
  year      = {2023},
  booktitle = {Interspeech 2023},
  pages     = {5281--5285},
  doi       = {10.21437/Interspeech.2023-1820},
  issn      = {2958-1796},
}

@inproceedings{kim25j_interspeech,
  title     = {Naturalness-Aware Curriculum Learning with Dynamic Temperature for Speech Deepfake Detection},
  author    = {Taewoo Kim and Guisik Kim and Choongsang Cho and Young Han Lee},
  year      = {2025},
  booktitle = {{Interspeech 2025}},
  pages     = {5318--5322},
  doi       = {10.21437/Interspeech.2025-717},
  issn      = {2958-1796},
}

@article{zhang2022partialspoof,
  title={The partialspoof database and countermeasures for the detection of short fake speech segments embedded in an utterance},
  author={Zhang, Lin and Wang, Xin and Cooper, Erica and Evans, Nicholas and Yamagishi, Junichi},
  journal={IEEE/ACM Transactions on Audio, Speech, and Language Processing},
  volume={31},
  pages={813--825},
  year={2022},
  publisher={IEEE}
}

@inproceedings{zhang2026compspoof,
  title={Compspoof: A dataset and joint learning framework for component-level audio anti-spoofing countermeasures},
  author={Zhang, Xueping and Wang, Yechen and Li, Linxi and Jin, Liwei and Li, Ming},
  booktitle={ICASSP 2026-2026 IEEE International Conference on Acoustics, Speech and Signal Processing (ICASSP)},
  pages={18067--18071},
  year={2026},
  organization={IEEE}
}

@inproceedings{zang2024singfake,
  title={Singfake: Singing voice deepfake detection},
  author={Zang, Yongyi and Zhang, You and Heydari, Mojtaba and Duan, Zhiyao},
  booktitle={ICASSP 2024-2024 IEEE International Conference on Acoustics, Speech and Signal Processing (ICASSP)},
  pages={12156--12160},
  year={2024},
  organization={IEEE}
}

@inproceedings{li2024cross,
  title={Cross-domain audio deepfake detection: Dataset and analysis},
  author={Li, Yuang and Zhang, Min and Ren, Mengxin and Qiao, Xiaosong and Ma, Miaomiao and Wei, Daimeng and Yang, Hao},
  booktitle={Proceedings of the 2024 Conference on Empirical Methods in Natural Language Processing},
  pages={4977--4983},
  year={2024}
}

@inproceedings{gu2025allm4add,
  title={Allm4add: Unlocking the capabilities of audio large language models for audio deepfake detection},
  author={Gu, Hao and Yi, Jiangyan and Wang, Chenglong and Tao, Jianhua and Lian, Zheng and He, Jiayi and Ren, Yong and Chen, Yujie and Wen, Zhengqi},
  booktitle={Proceedings of the 33rd ACM International Conference on Multimedia},
  pages={11736--11745},
  year={2025}
}

@article{guo2026towards,
  title={Towards Explicit Acoustic Evidence Perception in Audio LLMs for Speech Deepfake Detection},
  author={Guo, Xiaoxuan and Xie, Yuankun and Cheng, Haonan and Zhou, Jiayi and Liu, Jian and Huang, Hengyan and Ye, Long and Zhang, Qin},
  journal={arXiv preprint arXiv:2601.23066},
  year={2026}
}

@inproceedings{yamagishi21_asvspoof,
  title     = {ASVspoof 2021: accelerating progress in spoofed and deepfake speech detection},
  author    = {Junichi Yamagishi and Xin Wang and Massimiliano Todisco and Md Sahidullah and Jose Patino and Andreas Nautsch and Xuechen Liu and Kong Aik Lee and Tomi Kinnunen and Nicholas Evans and Héctor Delgado},
  year      = {2021},
  booktitle = {2021 Edition of the Automatic Speaker Verification and Spoofing Countermeasures Challenge},
  pages     = {47--54},
  doi       = {10.21437/ASVSPOOF.2021-8},
}

@article{schick2023toolformer,
  title={Toolformer: Language models can teach themselves to use tools},
  author={Schick, Timo and Dwivedi-Yu, Jane and Dess{\`\i}, Roberto and Raileanu, Roberta and Lomeli, Maria and Hambro, Eric and Zettlemoyer, Luke and Cancedda, Nicola and Scialom, Thomas},
  journal={Advances in neural information processing systems},
  volume={36},
  pages={68539--68551},
  year={2023}
}

@inproceedings{qin2024toolllm,
  title={Toolllm: Facilitating large language models to master 16000+ real-world apis},
  author={Yujia Qin and Shihao Liang and Yining Ye and Kunlun Zhu and Lan Yan and Yaxi Lu and Yankai Lin and Xin Cong and Xiangru Tang and Bill Qian and Sihan Zhao and Lauren Hong and Runchu Tian and Ruobing Xie and Jie Zhou and Mark Gerstein and dahai li and Zhiyuan Liu and Maosong Sun},
  booktitle={International Conference on Learning Representations},
  year={2024}
}

@article{qian2025toolrl,
  title={ToolRL: Reward is All Tool Learning Needs},
  author={Qian, Cheng and Acikgoz, Emre Can and He, Qi and Wang, Hongru and Chen, Xiusi and Hakkani-T{\"u}r, Dilek and Tur, Gokhan and Ji, Heng},
  journal={arXiv preprint arXiv:2504.13958},
  year={2025}
}

@inproceedings{babu22_interspeech,
  title     = {XLS-R: Self-supervised Cross-lingual Speech Representation Learning at Scale},
  author    = {Arun Babu and Changhan Wang and Andros Tjandra and Kushal Lakhotia and Qiantong Xu and Naman Goyal and Kritika Singh and Patrick {von Platen} and Yatharth Saraf and Juan Pino and Alexei Baevski and Alexis Conneau and Michael Auli},
  year      = {2022},
  booktitle = {{Interspeech 2022}},
  pages     = {2278--2282},
  doi       = {10.21437/Interspeech.2022-143},
  issn      = {2958-1796},
}

@article{rouditchenko2025omni,
  title={Omni-r1: Do you really need audio to fine-tune your audio llm?},
  author={Rouditchenko, Andrew and Bhati, Saurabhchand and Araujo, Edson and Thomas, Samuel and Kuehne, Hilde and Feris, Rogerio and Glass, James},
  journal={arXiv preprint arXiv:2505.09439},
  year={2025}
}

@article{xie2026interpretable,
  title={Interpretable All-Type Audio Deepfake Detection with Audio LLMs via Frequency-Time Reinforcement Learning},
  author={Xie, Yuankun and Guo, Xiaoxuan and Zhou, Jiayi and Wang, Tao and Liu, Jian and Fu, Ruibo and Wang, Xiaopeng and Cheng, Haonan and Ye, Long},
  journal={arXiv preprint arXiv:2601.02983},
  year={2026}
}

@article{nautsch2021asvspoof,
  title={ASVspoof 2019: Spoofing countermeasures for the detection of synthesized, converted and replayed speech},
  author={Nautsch, Andreas and Wang, Xin and Evans, Nicholas and Kinnunen, Tomi H and Vestman, Ville and Todisco, Massimiliano and Delgado, H{\'e}ctor and Sahidullah, Md and Yamagishi, Junichi and Lee, Kong Aik},
  journal={IEEE Transactions on Biometrics, Behavior, and Identity Science},
  volume={3},
  number={2},
  pages={252--265},
  year={2021},
  publisher={IEEE}
}

@article{comanducci2025fakemusiccaps,
  title={Fakemusiccaps: A dataset for detection and attribution of synthetic music generated via text-to-music models},
  author={Comanducci, Luca and Bestagini, Paolo and Tubaro, Stefano},
  journal={Journal of Imaging},
  volume={11},
  number={7},
  pages={242},
  year={2025},
  publisher={MDPI}
}

@article{agostinelli2023musiclm,
  title={Musiclm: Generating music from text},
  author={Andrea Agostinelli and Timo I. Denk and Zalán Borsos and Jesse Engel and Mauro Verzetti and Antoine Caillon and Qingqing Huang and Aren Jansen and Adam Roberts and Marco Tagliasacchi and Matt Sharifi and Neil Zeghidour and Christian Frank},
  journal={arXiv preprint arXiv:2301.11325},
  year={2023}
}

@inproceedings{yin25_interspeech,
  title     = {{EnvSDD: Benchmarking Environmental Sound Deepfake Detection}},
  author    = {Han Yin and Yang Xiao and Rohan Kumar Das and Jisheng Bai and Haohe Liu and Wenwu Wang and Mark D Plumbley},
  year      = {2025},
  booktitle = {{Interspeech 2025}},
  pages     = {201--205},
  doi       = {10.21437/Interspeech.2025-1143},
  issn      = {2958-1796},
}

@inproceedings{rouard2023hybrid,
  title={Hybrid transformers for music source separation},
  author={Rouard, Simon and Massa, Francisco and D{\'e}fossez, Alexandre},
  booktitle={ICASSP 2023-2023 IEEE International Conference on Acoustics, Speech and Signal Processing (ICASSP)},
  pages={1--5},
  year={2023},
  organization={IEEE}
}

@article{Qwen2.5-Omni,
  title={Qwen2.5-Omni Technical Report},
  author={Jin Xu and Zhifang Guo and Jinzheng He and Hangrui Hu and Ting He and Shuai Bai and Keqin Chen and Jialin Wang and Yang Fan and Kai Dang and Bin Zhang and Xiong Wang and Yunfei Chu and Junyang Lin},
  journal={arXiv preprint arXiv:2503.20215},
  year={2025}
}

@inproceedings{turpault2019sound,
  title={Sound event detection in domestic environments with weakly labeled data and soundscape synthesis},
  author={Turpault, Nicolas and Serizel, Romain and Shah, Ankit Parag and Salamon, Justin},
  booktitle={Workshop on Detection and Classification of Acoustic Scenes and Events},
  year={2019}
}

\appendix

\section{Mixed-Authenticity Benchmark Construction}
\label{sec:appendix_dataset}

This appendix provides additional details on the construction pipeline and statistics of the Mixed-Authenticity Audio Deepfake Detection Benchmark. We first describe how single-source audio clips are converted into structured component annotations, and then explain how these annotations are used to synthesize composite examples for the C1, C2, and C3 settings.

\begin{figure*}[t]
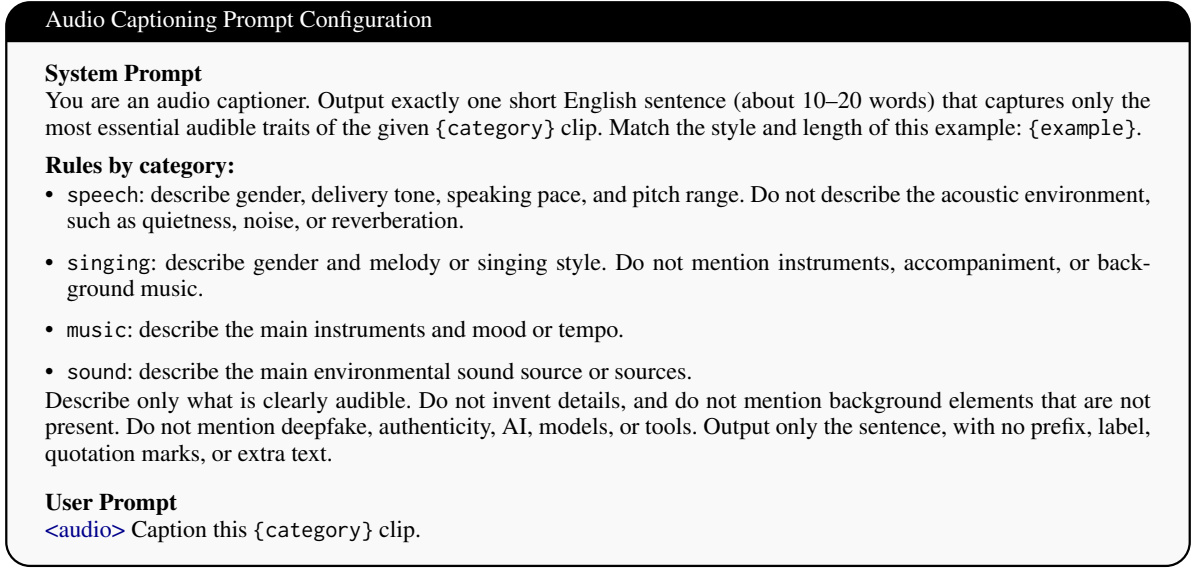

\centering
\begin{tcolorbox}[
colback=gray!5!white,
colframe=black,
title=Audio Captioning Prompt Configuration,
boxrule=0.3mm,
width=0.98\linewidth,
arc=3mm,
auto outer arc=true,
fontupper=\footnotesize,
fonttitle=\small
]

\textbf{System Prompt} \\
You are an audio captioner. Output exactly one short English sentence
(about 10--20 words) that captures only the most essential audible traits
of the given \texttt{\{category\}} clip. Match the style and length of
this example: \texttt{\{example\}}.

\vspace{1.5mm}
\textbf{Rules by category:}
\begin{itemize}[leftmargin=*,itemsep=1pt,topsep=1pt]
    \item \texttt{speech}: describe gender, delivery tone, speaking pace, and pitch range. Do not describe the acoustic environment, such as quietness, noise, or reverberation.
    \item \texttt{singing}: describe gender and melody or singing style. Do not mention instruments, accompaniment, or background music.
    \item \texttt{music}: describe the main instruments and mood or tempo.
    \item \texttt{sound}: describe the main environmental sound source or sources.
\end{itemize}

Describe only what is clearly audible. Do not invent details, and do not
mention background elements that are not present. Do not mention
deepfake, authenticity, AI, models, or tools. Output only the sentence,
with no prefix, label, quotation marks, or extra text.

\vspace{3.0mm}
\textbf{User Prompt} \\
\textcolor{deepblue}{\textless{}audio\textgreater{}} Caption this \texttt{\{category\}} clip.

\end{tcolorbox}
\caption{Prompt configuration used for Qwen3-Omni-30B-A3B-Captioner-based single-source audio caption generation. The prompt constrains captions to observable acoustic properties and excludes authenticity-related information to avoid label leakage.}
\label{fig:caption_prompt}
\end{figure*}

\subsection{Audio Caption and Component Annotation}
\label{sec:appendix_caption_annotation}

Before constructing composite mixtures, we annotate each single-source audio clip with structured component-level metadata. Each component is associated with a content type, a support region, and an authenticity label. The content type is selected from speech, singing, music, and environmental sound. The support region specifies the temporal span in which the component is active, and the authenticity label indicates whether the component is real or fake.

We also generate a concise single-source audio caption for each component using Qwen3-Omni-30B-A3B-Captioner\footnote{\url{https://huggingface.co/Qwen/Qwen3-Omni-30B-A3B-Captioner}}. The caption describes only observable acoustic properties, such as vocal delivery, singing style, instrumentation, or environmental sound sources, depending on the component type. Importantly, the captioning prompt explicitly prevents the model from mentioning authenticity-related information, including whether the audio is real, fake, synthetic, AI-generated, or produced by a model. The detailed prompt configuration and category-specific rules are provided in Figure~\ref{fig:caption_prompt}. This design avoids label leakage while preserving the acoustic information needed for trajectory construction.

After composite mixtures are constructed, the captions are combined with the component metadata to form the \texttt{<audio\_understanding>} block of the supervised ToolDF trajectory. Thus, the model is supervised not only on the final clip-level authenticity label, but also on the intermediate acoustic structure needed for tool selection and evidence aggregation.

\paragraph{Boundary Annotation.}
For vocal domains, including speech and singing, we estimate active regions using WebRTC-VAD\footnote{\url{https://github.com/wiseman/py-webrtcvad}}. This excludes prolonged silence and yields localized support regions for vocal components. For non-vocal domains, including music and environmental sound, we use the full clip duration as the valid support region, since these signals typically function as continuous background components in the constructed mixtures.

\begin{table*}[t]
\centering
\small
\renewcommand{\arraystretch}{1.15} 
\begin{tabular}{llcccc}
\toprule
\textbf{Split} & \textbf{Acoustic Configuration} & \textbf{train} ($R$/$F$) & \textbf{dev} ($R$/$F$) & \textbf{eval} ($R$/$F$) & \textbf{Total} \\ 
\midrule
\textbf{C1} & Speech $\rightarrow$ Singing & 645 / 12,045 & 637 / 11,785 & 1,838 / 33,778 & 60,728 \\
(Temporal) & Singing $\rightarrow$ Speech & 645 / 12,045 & 637 / 11,785 & 1,839 / 33,781 & 60,732 \\ 
\cmidrule{2-6} 
 & \textbf{C1 Subtotal} & \textbf{1,290 / 24,090} & \textbf{1,274 / 23,570} & \textbf{3,677 / 67,559} & \textbf{121,460} \\ 
\midrule
\textbf{C2} & Speech + Music & 159 / 3,011 & 170 / 2,155 & 170 / 2,944 & 8,609 \\
(Overlap) & Singing + Music & 657 / 9,624 & 361 / 3,734 & 354 / 3,803 & 18,533 \\
 & Speech + Environmental Sound & 1,155 / 21,055 & 1,108 / 12,952 & 867 / 16,367 & 53,504 \\
 & Singing + Environmental Sound & 5,403 / 68,719 & 2,867 / 22,783 & 1,630 / 20,904 & 122,306 \\ 
\cmidrule{2-6} 
 & \textbf{C2 Subtotal} & \textbf{7,374 / 102,409} & \textbf{4,506 / 41,624} & \textbf{3,021 / 44,018} & \textbf{202,952} \\ 
\midrule
\textbf{C3} & (Speech $\rightarrow$ Singing) + Music & 33 / 1,385 & 40 / 1,157 & 154 / 825 & 3,594 \\
(Hybrid) & (Singing $\rightarrow$ Speech) + Music & 30 / 1,461 & 35 / 1,187 & 177 / 808 & 3,698 \\
 & (Speech $\rightarrow$ Singing) + EnvSound & 293 / 10,979 & 291 / 7,524 & 758 / 4,303 & 24,148 \\
 & (Singing $\rightarrow$ Speech) + EnvSound & 289 / 10,910 & 271 / 7,519 & 749 / 4,310 & 24,048 \\ 
\cmidrule{2-6} 
 & \textbf{C3 Subtotal} & \textbf{645 / 24,735} & \textbf{637 / 17,387} & \textbf{1,838 / 10,246} & \textbf{55,488} \\ 
\midrule[\heavyrulewidth] 
\textbf{Total} & \textbf{Combined Composite Pool} & \textbf{9,309 / 151,234} & \textbf{6,417 / 82,581} & \textbf{8,536 / 121,823} & \textbf{379,900} \\ 
\bottomrule
\end{tabular}
\caption{Detailed statistics of the composite-type configurations in the mixed-authenticity audio deepfake detection benchmark. C1 denotes temporal speech--singing transitions, C2 denotes acoustic overlaps between vocal and non-vocal components, and C3 denotes hybrid mixtures combining both.}
\label{tab:detailed_stats_appendix}
\end{table*}

\subsection{Composite Mixture Construction}
\label{sec:appendix_composite_construction}

Using the annotated single-source component pools, we construct three types of composite mixed-authenticity examples: temporal transitions, acoustic overlaps, and hybrid mixtures. Each composite example is created by combining components from different acoustic domains while preserving their component-level support regions and authenticity labels. The clip-level label is then assigned according to the early-fail rule: a composite clip is labeled fake if at least one of its constituent components is fake, and real otherwise.

\paragraph{Temporal Transition (C1).}
The C1 setting models temporal transitions between two vocal domains. We construct each example by sequentially concatenating a speech clip and a singing clip. Both transition orders are considered: Speech $\rightarrow$ Singing and Singing $\rightarrow$ Speech. The support region of each component is updated according to its position in the concatenated sequence, so that the resulting trajectory can identify which temporal segment corresponds to speech and which corresponds to singing.

To cover different mixed-authenticity patterns, we construct four
authenticity combinations: real-to-real ($R \rightarrow R$),
real-to-fake ($R \rightarrow F$), fake-to-real ($F \rightarrow R$), and fake-to-fake ($F \rightarrow F$). The first three combinations are sampled with the same budget to ensure balanced coverage of real and partially fake transitions. The fake-to-fake subset is then filled using the remaining available fake source pools. This design allows C1 to evaluate whether a model can detect localized manipulations across temporal changes in vocal content.

\paragraph{Acoustic Overlap (C2).}
The C2 setting models overlapping foreground and background sources. We mix a foreground vocal component with a non-vocal background component. The foreground component is either speech or singing, and the background component is either music or environmental sound. This results in four sub-configurations: Speech + Music, Speech + Environmental Sound, Singing + Music, and Singing + Environmental Sound.

For each mixture, the vocal component retains its localized support region, while the background component is treated as active over the full mixture duration. The resulting example therefore contains source-level overlap between a foreground vocal stream and a background non-vocal stream. The number of examples in each sub-configuration follows the available source-pool sizes, which naturally yields larger subsets for singing and environmental sound.

\begin{figure*}[ht]
\centering
\begin{tcolorbox}[
colback=gray!5!white,
colframe=black,
title=ToolDF Prompt Configuration,
boxrule=0.3mm,
width=0.98\linewidth,
arc=3mm,
auto outer arc=true,
fontupper=\footnotesize,
fonttitle=\small
]

\textbf{System Prompt} \\
You are an expert in audio deepfake detection. You MUST use the provided tools to judge authenticity — never decide by yourself, and never fabricate tool results. \\

Within \texttt{<think>}, write \texttt{<audio\_understanding>} (segment-by-segment analysis with timestamps) and \texttt{<plan>} (which detectors to call on which segments, and whether \texttt{source\_separator} is needed). \\

After tool results return, write \texttt{<description>} summarizing each tool's prediction and score, then \texttt{<answer>} with a single word: \texttt{real} or \texttt{fake}. Any fake detection means the overall answer is \texttt{fake}.

\vspace{3.0mm}
\textbf{User Prompt} \\
\textcolor{deepblue}{\textless{}audio\textgreater{}} Is this audio real or fake?

\end{tcolorbox}
\caption{System prompt used during supervised fine-tuning and inference to enforce the tool-integrated reasoning protocol for the ToolDF orchestrator.}
\label{fig:system_prompt_appendix}
\end{figure*}

\paragraph{Hybrid Mixture (C3).}
The C3 setting combines the two previous composition types. We first construct a C1-style temporal transition between speech and singing, and then overlay the resulting vocal sequence with a non-vocal background component, either music or environmental sound. This produces hybrid mixtures that contain both temporal changes in the foreground vocal content and acoustic overlap with a background source.

As in C2, the background component is treated as active over the full mixture duration, while the speech and singing components retain their updated temporal support regions from the transition sequence. C3 is therefore the most structurally complex setting, requiring the model to reason over both temporal segmentation and source-level separation before aggregating component-level authenticity evidence.

\subsection{Dataset Statistics}
\label{sec:appendix_statistics}
The complete benchmark contains 973,649 examples, including 379,900 composite examples from the C1, C2, and C3 settings. Table~\ref{tab:detailed_stats_appendix} reports the detailed composite-type statistics across train, development, and evaluation splits. The composite pool is fake-heavy because a clip is labeled fake whenever at least one constituent component is fake; the higher proportion of real samples in the C3 evaluation split results from basename-level resplitting of FakeMusicCaps samples to prevent background-track leakage.

\section{System Prompt for ToolDF}
\label{sec:appendix_system_prompt}

To ensure the reproducibility of the Tool-Integrated Reasoning (TIR) trajectory, we provide the exact system prompt used by the ALLM orchestrator during both supervised fine-tuning (SFT) and inference. As specified in Section~\ref{sec:sft}, this prompt enforces the structured generation format and prevents the model from fabricating tool results or deviating from the prescribed reasoning protocol.

\end{document}